# Thermally Robust and Strongly Pressure-Responsive NIR Luminescent Manometer Based on $Cr^{3+}$-$Ni^{2+}$ Emission

**Weijia Xie [a #], Maja Szymczak[b#], Tongxi Luo [a], Chan Wang [a *], Lukasz Marciniak [b *], Xinguo Zhang [a*]**

[a] NMPA Key Laboratory for Research and Evaluation of Drug Metabolism & Guangdong Provincial Key Laboratory of New Drug Screening & Guangdong-Hongkong-Macao Joint Laboratory for New Drug Screening, School of Pharmaceutical Sciences, Southern Medical University, Guangzhou 510515, Guangdong, China

[b] Institute of Low Temperature and Structure Research Polish Academy of Sciences, 50-422 Wroclaw, Poland

[#] These authors contributed equally to this work.

Corresponding authors: wangchan0308@163.com (C. Wang), l.marciniak@intibs.pl (L. Marciniak), mpcc1@qq.com (X. Zhang)

**Abstract:**

The increasing demand for highly sensitive luminescent manometers capable of precise pressure determination, particularly those operating in the near-infrared (NIR) spectral range, motivates the continuous search for new pressure-responsive luminescent materials. In this work, we introduce $NaLu_2Ga_3Ge_2O_{12}:Cr^{3+}$, $Ni^{2+}$ as a highly sensitive NIR luminescent manometer in which pressure not only induces pronounced spectral shifts of the $Cr^{3+}$ and $Ni^{2+}$ emission bands but also strongly modulates their relative luminescence intensities. Remarkably, the $^3T_2(^3F)\rightarrow{}^3A_2(^3F)$ emission band of $Ni^{2+}$ exhibits a record-high absolute pressure sensitivity of $S_A$ = 34 nm GPa$^{-1}$. Furthermore, by appropriately selecting narrow spectral ranges within the pressure-responsive emission bands, a ratiometric NIR luminescent manometer was developed, providing a maximum relative pressure sensitivity of $S_R$ = 158% GPa$^{-1}$. Most importantly, the proposed ratiometric readout exhibits a thermal-invariance manometric factor (*TIMF*) of 19,900 K GPa$^{-1}$, which, to the best of our knowledge, is the highest value reported to date for a luminescent manometer. The combination of exceptionally high-pressure sensitivity, NIR operation, and remarkable resistance to thermal interference establishes $NaLu_2Ga_3Ge_2O_{12}$:$Cr^{3+}$, $Ni^{2+}$ as a highly promising luminescent pressure sensor for reliable pressure determination under conditions involving simultaneous temperature variations.

## Introduction

Given that pressure, alongside temperature, is one of the most important thermodynamic parameters from both fundamental and technological perspectives, the widespread development and application of various pressure-sensing techniques is not surprising[1,2]. Among them, an particularly important approach exploits pressure-induced changes in the spectroscopic properties of phosphor materials, forming the basis of luminescence manometry[3–5]. This technique enables remote optical pressure determination with high precision and is particularly attractive under high-pressure conditions, typically above 1 GPa. The benchmark material for luminescence manometry is $Al_2O_3:Cr^{3+}$ (ruby), in which the spectral position of the $Cr^{3+}$ $^2E \rightarrow ^4A_2$ emission line undergoes a monotonic redshift with increasing pressure, primarily due to the pressure-induced modification of the nephelauxetic effect[6–12]. The preservation of this monotonic response up to pressures approaching 150 GPa, together with the excellent chemical and mechanical stability of ruby, has established it as the most widely employed luminescent pressure calibrant. Nevertheless, ruby suffers from several important limitations, including its relatively low-pressure sensitivity and the simultaneous dependence of the emission-line position on both pressure and temperature. Furthermore, because pressure determination relies primarily on monitoring the spectral position of the emission line, ruby is not ideally suited for rapid visualization of spatially nonuniform pressure distributions[3,13]. Mapping such distributions would require point-by-point spectral acquisition across the region of interest, substantially limiting the achievable acquisition rate. These limitations have stimulated intensive efforts toward the development of alternative luminescent manometers[14–19]. In many practical environments, visible luminescence may additionally undergo substantial absorption or scattering by the medium surrounding the phosphor[20,21]. Consequently, luminescent manometers operating in the near-infrared (NIR) spectral range are particularly attractive, as optical absorption and scattering can be reduced within appropriate NIR spectral windows. In this context, considerable attention has been devoted to $Cr^{3+}$-doped materials, particularly those exhibiting the broadband $^4T_2 \rightarrow ^4A_2$ emission[22–26]. The energy of the $^4T_2$ state is strongly dependent on the crystal-field strength and therefore responds sensitively to pressure-induced changes in the local coordination environment of $Cr^{3+}$. However, the spectral position of this emission is typically restricted to approximately 800-1100 nm, limiting access to longer-wavelength NIR regions.

To address these limitations, in the present work we demonstrate the multiple advantages arising from the co-doping of $NaLu_2Ga_3Ge_2O_{12}$ with $Cr^{3+}$ and $Ni^{2+}$ ions. The unique spectroscopic interplay between these ions not only substantially enhances the $Ni^{2+}$ luminescence intensity through $Cr^{3+}$→$Ni^{2+}$ energy transfer, but also enables multimodal optical pressure readout. Detailed analysis reveals that the pressure-induced spectral shift of the $Ni^{2+}$ emission in $NaLu_2Ga_3Ge_2O_{12}$:$Cr^{3+}$, $Ni^{2+}$ provides an exceptionally high relative pressure sensitivity, exceeding that reported for previously investigated luminescent manometers. Moreover, the combination of the pressure-induced spectral shifts of the $Cr^{3+}$ and $Ni^{2+}$ emission bands with the pressure-dependent efficiency of $Cr^{3+}$→$Ni^{2+}$ energy transfer enables a highly sensitive ratiometric pressure readout with $S_{Rmax}$ = 158% $GPa^{-1}$. Most importantly, the appropriately defined ratiometric parameter exhibits a pronounced response to pressure while remaining nearly invariant to temperature. This exceptionally high pressure selectivity results in a *TIMF* of 19,900 K $GPa^{-1}$, representing, to the best of our knowledge, the highest value reported for a luminescent manometer to date. Thus, the $Cr^{3+}$,$Ni^{2+}$ co-doped system simultaneously combines NIR emission, sensitized $Ni^{2+}$ luminescence, multimodal pressure readout, exceptionally high-pressure sensitivity, and strongly suppressed thermal cross-sensitivity, providing a promising platform for the development of highly sensitive and temperature-invariant optical pressure sensors.

**Experimental Section**

*Synthesis*

$NaLu_2Ga_{3-x-y}Ge_2O_{12}$:$x$$Ni^{2+}$, $y$$Cr^{3+}$ ($x$=0~0.13, $y$=0~0.30) compounds were synthesized via a high-temperature solid-state method. The stoichiometric amounts of $Na_2CO_3$, $Lu_2O_3$, $Ga_2O_3$, $GeO_2$, NiO, $Cr_2O_3$ powders were weighed and thoroughly mixed in an agate mortar for 10 min, with 1 wt % $Li_2CO_3$ added as flux. Subsequently, the mixtures were sintered at 1350°C for 5 h in air. After cooling to room temperature, the samples were reground for 10 min to obtain the final phosphor.

*Characterization*

Structural characterization of the $Ni^{2+}$- doped and $Cr^{3+}$-doped $NaLu_2Ga_3Ge_2O_{12}$ phosphor was performed using X-ray diffraction (XRD, Bruker D8) with Cu-Kα radiation (40 kV,40 mA). Chemical composition analysis was conducted through X-ray photoelectron spectroscopy (XPS) employing a Thermo SCIENTIFIC Nexsa. The surface morphology and dimensional characteristics of the synthesized materials were analyzed by Scanning electron microscopy (SEM) coupled with energy-dispersive X-ray spectroscopy (EDS) mapping using a TESCAN MIRA LMS instrument. Diffuse reflectance characteristics were quantified in the region (200-1000 nm) employing a Shimadzu UV-3600 spectrophotometer. The optical properties were systematically investigated with an Edinburgh FLS980 spectrophotometer, acquiring photoluminescence excitation (PLE) spectra, emission (PL) spectra, and luminescence decay profiles. The Edinburgh FLS-1000 fluorescence spectrophotometer with a 450 W xenon lamp as the excitation source, was used to monitor temperature- and pressure-dependent PL spectra.

A 445 nm laser diode was used to excite the samples during both temperature- and pressure-dependent luminescence measurements, including emission spectra and decay kinetics. Temperature-dependent experiments were carried out using a THMS 600 heating-cooling stage (Linkam), providing a temperature stability and set-point resolution of 0.1 K. At each temperature, the sample was allowed to equilibrate for 2 min before data acquisition.

High-pressure luminescence measurements were performed in a nitrogen gas membrane-driven diamond anvil cell (Diacell μScopeDAC-RT(G), Almax easyLab), with pressure controlled by a Druck PACE 5000 system. The DAC was equipped with ultra-low-fluorescence type IIa diamond anvils with 0.4 mm culets. A stainless-steel gasket, initially 250 μm thick and 10 mm in diameter, was pre-indented, and a 140 μm hole was drilled in its center to serve as the sample chamber. The investigated material and $SrB_2O_4:Sm^{2+}$ pressure calibrant were placed inside the chamber[27]. A 4:1 (v/v) methanol-ethanol mixture served as the pressure-transmitting medium, providing quasi-hydrostatic conditions over the investigated pressure range.

The experimental luminescence decay profiles were analyzed using a biexponential decay function:

$$I(t) = I_0 + A_1 \exp\left(\frac{t}{\tau_1}\right) + A_2 \exp\left(\frac{t}{\tau_2}\right) \quad (1)$$

$$\tau = \frac{A_1\tau_1^2 + A_2\tau_2^2}{A_1\tau_1 + A_2\tau_2} \quad (2)$$

In this model, $\tau_1$ and $\tau_2$ correspond to the decay times, whereas $A_1$ and $A_2$ are the associated pre-exponential coefficients.

**Results and Discussions**

The structure of $NaLu_2Ga_3Ge_2O_{12}$ consists of three distinct coordination polyhedral, i.e. 6-fold coordinated $Ga^{3+}$, 8-fold coordinated $Lu^{3+}$ and 4-fold-coordinated $Ga^{3+}$ (Figure 1a)[28–31]. Among them, $Na^{+}$ and $Lu^{3+}$ atoms occupy the same crystallographic site. $Cr^{3+}$ ions preferentially adopt octahedral coordination; therefore, when $NaLu_2Ga_3Ge_2O_{12}$ is doped with $Cr^{3+}$, the 6-fold coordinated $Ga^{3+}$ site represents the most probable crystallographic position occupied by these ions. In tetrahedrally coordinated sites, chromium is typically stabilized in the $Cr^{4+}$ oxidation state[32–36]. Importantly, $Cr^{4+}$ exhibits spectroscopic properties distinctly different from those of $Cr^{3+}$, enabling an unambiguous distinction between these two oxidation states. In the case of $Ni^{2+}$ ions, occupation of both 6-fold and 4-fold coordinated crystallographic sites has been reported in the literature, although 6-fold coordination is generally considered thermodynamically more favorable[37–39]. However, because of the charge mismatch between $Ni^{2+}$ dopant ions and the $Ga^{3+}$ host cations, such substitution may require charge compensation and consequently promote the formation of structural defects. For this reason, very low dopant concentrations were employed in the present study to minimize any dopant-induced perturbation of the host crystal structure. Structural analysis of the materials synthesized with different dopant concentrations revealed excellent agreement between their room-temperature powder X-ray diffraction patterns and the corresponding reference pattern, confirming the phase purity of all synthesized samples and indicating the absence of detectable secondary phases (Figure 1b). Furthermore, Rietveld refinement of the X-ray diffraction data supported the proposed substitution of the dopant ions at the $Ga^{3+}$ sites (the refinement results are listed in Tables S1 and S2) (Figure 1c and S1). The doping of $Cr^{3+}$ with adopted concentrations does not brings significant change on the peak position of the XRD patterns, as evidenced by matching their diffraction peaks with the calculated $NaLu_2Ga_3Ge_2O_{12}$: $Cr^{3+}$, $Ni^{2+}$ pattern (bottom) (Figure 1b). Based on the full-range XPS spectra of $NaLu_2Ga_3Ge_2O_{12}$: $Cr^{3+}$, $Ni^{2+}$ samples the oxidation states of Ni and Cr in the sample were analyzed (the spectra were calibrated according to the binding energy of carbon peaks) (Figure 1d). It can be preliminarily concluded that nickel exists predominantly as $Ni^{2+}$ and chromium as $Cr^{3+}$ in the as-prepared samples (Figure 1d-f). On the basis of the high-resolution Ni2p XPS spectra and previously reported theoretical literatures[40,41], Ni is mainly stabilized in the divalent state ($Ni^{2+}$) (Figure 1e). The binding energy position and peak profile of the dominant Ni $2p_{3/2}$ peak, together with the distinct satellite peaks at the high-binding-energy side, match well with the typical XPS fingerprints of $Ni^{2+}$ species. Further confirmation from the electronic configuration and orbital splitting rules of Ni2p demonstrates that nickel is dominated by the $Ni^{2+}$ species with negligible other nickel valence states. Similarly, the high-resolution Cr 2p XPS spectrum confirms that chromium is predominantly present in the trivalent state ($Cr^{3+}$) (Figure 1f). The Cr $2p_{3/2}$ peak located at approximately 577 eV is consistent with the characteristic binding-energy range of $Cr^{3+}$ species, with no obvious contribution from higher chromium oxidation states. These results solidly verify the successful incorporation of $Ni^{2+}$ and $Cr^{3+}$ into the octahedral Ga crystallographic sites of the $NaLu_2Ga_3Ge_2O_{12}$:$0.05Ni^{2+}$, $Cr^{3+}$ host lattice.

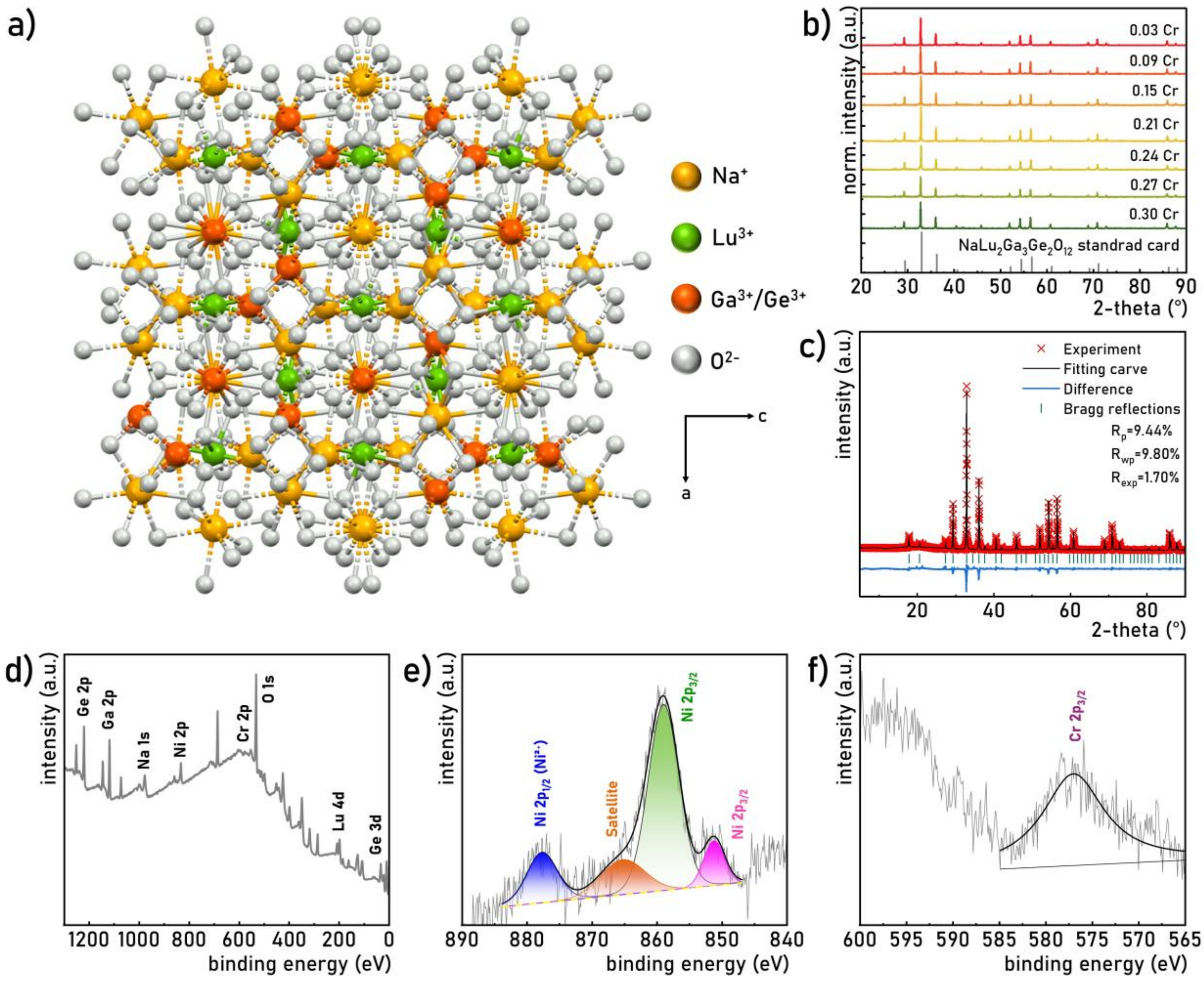

**Figure 1**. Crystal structure of $NaLu_2Ga_3Ge_2O_{12}$ -a); XRD patterns of $NaLu_2Ga_3Ge_2O_{12}$:0.05$Ni^{2+}$, $Cr^{3+}$ with different concentrations of $Cr^{3+}$ ions -b); the Rietveld refinement of $NaLu_2Ga_3Ge_2O_{12}$:0.05$Ni^{2+}$, 0.09$Cr^{3+}$ phosphors -c). the full XPS spectrum of $NaLu_2Ga_3Ge_2O_{12}$:0.13%$Ni^{2+}$, 0.24%$Cr^{3+}$ -d); Ni2p -e) and Cr2p -f) spectra with the fitting result.

The spectroscopic properties of $Cr^{3+}$ and $Ni^{2+}$ ions, as well as their response to externally applied pressure, can be rationalized using the Tanabe-Sugano diagrams for ions with $3d^3$ ($Cr^{3+}$) and $3d^8$ ($Ni^{2+}$) electronic configurations (Figure 2a). In the case of $Cr^{3+}$, depending on the crystal-field strength (CFS), either the $^2E$ or $^4T_2$ state constitutes the lowest-energy excited state.

Since the energy of the $^4T_2$ state is strongly dependent on CFS, variations in this parameter substantially modify the spectroscopic properties of $Cr^{3+}$ ions[42–44]. In materials characterized by a strong crystal field ($Dq/B > 2.2$), the emission spectrum is dominated by the spectrally narrow $^2E \rightarrow {}^4A_2$ transition. When the crystal-field strength decreases to the intermediate regime (approximately $1.9 < Dq/B < 2.2$), the energies of the $^2E$ and $^4T_2$ states become comparable, and both the narrow $^2E \rightarrow {}^4A_2$ emission and the broadband $^4T_2 \rightarrow {}^4A_2$ emission can be observed. A further decrease in CFS stabilizes the $^4T_2$ state below $^2E$, resulting in emission dominated by the broadband associated with the $^4T_2 \rightarrow {}^4A_2$ transition. Importantly, the crystal-field strength can be modified not only through changes in the chemical composition of the host material but also dynamically by applying external pressure. In strong CFS $Cr^{3+}$-doped phosphors, compression modifies the $Cr^{3+}$-ligand bond lengths and the nephelauxetic effect, resulting in a relatively modest spectral shift of the $^2E \rightarrow {}^4A_2$ emission line. In weak CFS systems, in contrast, pressure-induced changes in crystal-field strength strongly affect the energy of the $^4T_2$ state and consequently produce a pronounced spectral shift of the broadband $^4T_2 \rightarrow {}^4A_2$ emission. For $Ni^{2+}$ ions, a relatively weak crystal-field regime is commonly encountered, and their luminescence is therefore dominated by the broadband $^3T_2(^3F) \rightarrow {}^3A_2(^3F)$ transition. Analogously to the behavior of $Cr^{3+}$, increasing pressure strengthens the crystal field experienced by $Ni^{2+}$ ions, leading to a blueshift of the $Ni^{2+}$ emission band. Analysis of the excitation spectrum of $NaLu_2Ga_3Ge_2O_{12}$:$Ni^{2+}$ reveals a band centered at approximately 400 nm and a broad band extending from approximately 670 to 760 nm, which can be assigned to the $^3A_2(^3F) \rightarrow {}^3T_1(^3P)$ and $^3A_2(^3F) \rightarrow {}^3T_1(^3F)$ transitions of $Ni^{2+}$, respectively (Figure 2b). Upon excitation at $\lambda = 445$ nm, a broad NIR emission band centered at approximately 1450 nm is observed and attributed to the $^3T_2(^3F)\rightarrow^3A_2(^3F)$ transition. In the excitation spectrum of $NaLu_2Ga_3Ge_2O_{12}$: $Cr^{3+}$ characteristic bands are observed in the ultraviolet (~290 nm), blue (~450 nm), and red (~630 nm) spectral regions. These features are associated with host absorption and the $^4A_2(^4F)\rightarrow^4T_1(^4P)$, $^4A_2(^4F)\rightarrow^4T_1(^4F)$, and $^4A_2(^4F)\rightarrow^4T_2(^4F)$ transitions of $Cr^{3+}$, respectively. Owing to the relatively low absorption cross-section of $Ni^{2+}$, the excitation spectra of $NaLu_2Ga_3Ge_2O_{12}$:$Cr^{3+}$, $Ni^{2+}$ are dominated by the $Cr^{3+}$ absorption bands, irrespective of whether the $Cr^{3+}$ or $Ni^{2+}$ emission is monitored. Moreover, excitation within the $Cr^{3+}$ absorption range results in the simultaneous observation of both broadband $Cr^{3+}$ and $Ni^{2+}$ emission. This behavior provides clear evidence that population of the $Ni^{2+}$ $^3T_2$ excited state in the co-doped system occurs predominantly through $Cr^{3+}\rightarrow Ni^{2+}$ energy transfer. Since the probability of energy transfer is strongly dependent on the donor-acceptor separation, which can be controlled by varying the dopant concentrations, the influence of $Ni^{2+}$ concentration on the luminescence properties of $NaLu_2Ga_3Ge_2O_{12}$:$Cr^{3+}$, $Ni^{2+}$ was systematically investigated (Figure 2c, Figure S2-S3). The obtained spectra clearly demonstrate that increasing the $Ni^{2+}$ concentration leads to a progressive decrease in the $Cr^{3+}$ emission intensity relative to that of $Ni^{2+}$, providing further evidence for increasingly efficient $Cr^{3+}\rightarrow Ni^{2+}$ energy transfer. To quantify this effect, the luminescence intensity ratio $LIR_1$ was defined according to the following equation:

$$LIR_1 = \frac{\int_{1300\text{nm}}^{1600nm} {}^3T_2 \rightarrow {}^3A_2 d\lambda[Ni^{2+}]}{\int_{720\text{nm}}^{1100nm} {}^4T_2 \rightarrow {}^4A_2 d\lambda[Cr^{3+}]} \tag{3}$$

$LIR_1$ increases approximately linearly with increasing $Ni^{2+}$ concentration, confirming the systematic redistribution of the excitation energy from $Cr^{3+}$ toward $Ni^{2+}$ emission (Figure 2b).

The $Ni^{2+}$ concentration also has a pronounced influence on the luminescence quantum yield (QY) (Figure 2e). Initially, increasing the $Ni^{2+}$ concentration results in a progressive enhancement of QY, reaching a maximum value of 16.2% for the sample containing 0.003% $Ni^{2+}$. Further increasing the $Ni^{2+}$ concentration leads to a gradual decrease in QY. The decrease in the QY observed for higher concentrations of $Ni^{2+}$ ions can be explained through the $Cr^{3+}\rightarrow Ni^{2+}$ energy transfer which leads to the quenching of $Cr^{3+}$ emission band. To directly evaluate the evolution of the $Cr^{3+}\rightarrow Ni^{2+}$ energy-transfer process with $Ni^{2+}$ concentration, the $Cr^{3+}$ luminescence decay kinetics were systematically investigated (Figure 2f). Analysis of the decay curves reveals a pronounced shortening of the $Cr^{3+}$ excited-state lifetime from approximately $\tau$ = 34 μs in the absence of $Ni^{2+}$ to approximately $\tau$ = 6 μs for the sample containing 0.14% $Ni^{2+}$ (Figure 2g). The most pronounced reduction occurs at $Ni^{2+}$ concentrations below approximately 0.05%, indicating a rapid increase in the efficiency of the additional $Cr^{3+}$ depopulation pathway introduced by $Ni^{2+}$ ions. Based on the $Cr^{3+}$ excited-state lifetimes measured in the absence ($\tau_0$) and presence ($\tau$) of $Ni^{2+}$, the $Cr^{3+}\rightarrow Ni^{2+}$ energy-transfer efficiency was determined according to:

$$\eta = \left(1 - \frac{\tau}{\tau_0}\right) \cdot 100\% \qquad (4)$$

The calculated energy-transfer efficiency reaches approximately 24% already at a $Ni^{2+}$ concentration of 0.003% and increases systematically with increasing $Ni^{2+}$ content, reaching approximately 79% for 0.15% $Ni^{2+}$ (Figure 2g). This value exceeds those of most previously reported $Cr^{3+}$, $Ni^{2+}$ co-doped counterparts including $Sr_2GaTaO_6$:0.02$Cr^{3+}$, 0.04$Ni^{2+}$ (70%)[45] and $Li_2ZnSn_3O_8$:0.03$Cr^{3+}$, 0.03$Ni^{2+}$ (45.5%)[46], demonstrating the outstanding energy-transfer capability of the $NaLu_2Ga_3Ge_2O_{12}:Cr^{3+}$, $Ni^{2+}$ phosphor. These results provide direct spectroscopic evidence for highly efficient and concentration-dependent $Cr^{3+}\rightarrow Ni^{2+}$ energy transfer in $NaLu_2Ga_3Ge_2O_{12}:Cr^{3+}$, $Ni^{2+}$ and demonstrate that the relative contributions of $Cr^{3+}$ and $Ni^{2+}$ luminescence can be effectively controlled through the dopant concentration.

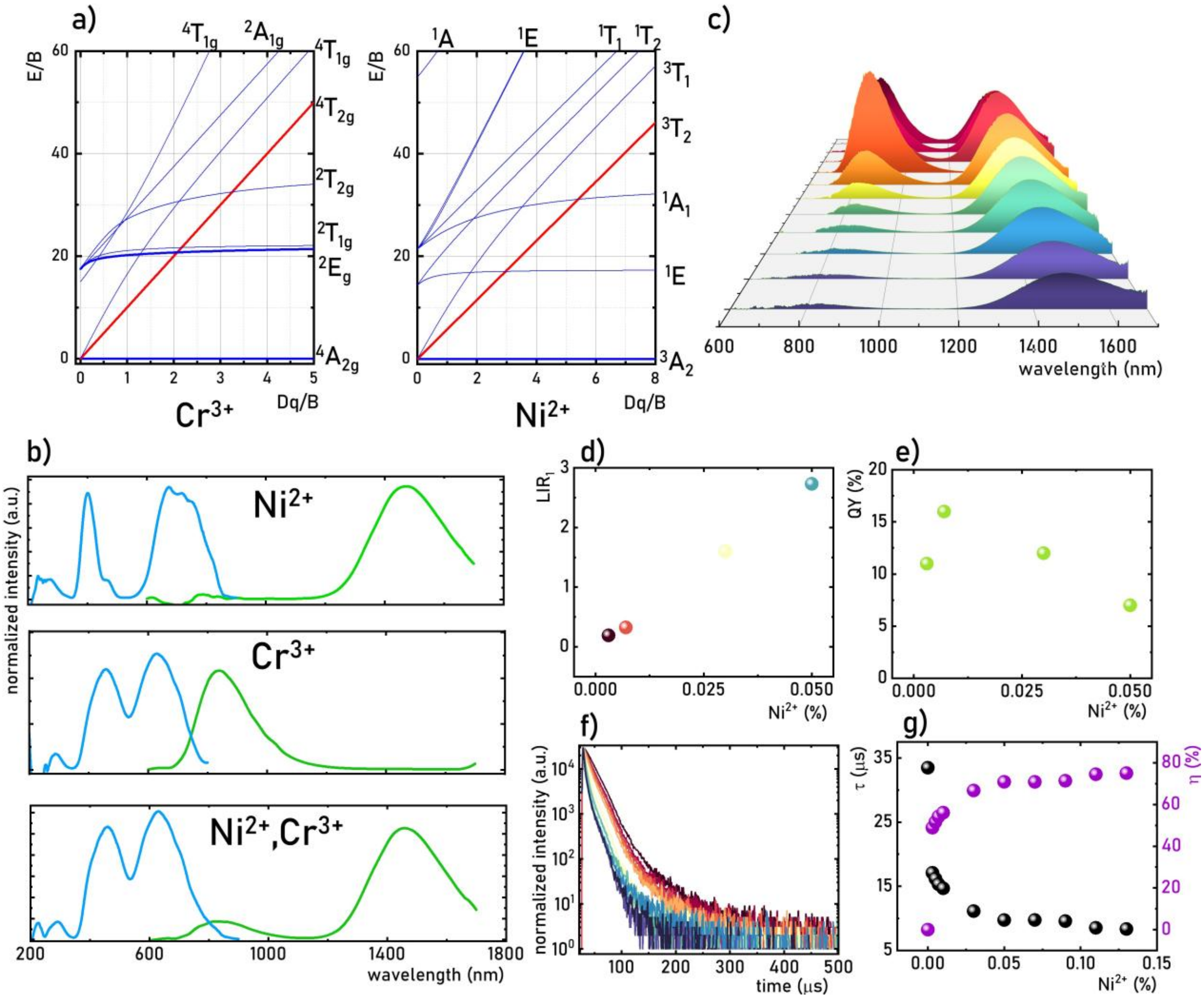

**Figure 2**. Simplified Tanabe-Sugano diagram for $Cr^{3+}$ and $Ni^{2+}$ ions -a), comparison of the emission and excitation spectra of $NaLu_2Ga_3Ge_2O_{12}:Ni^{2+}$ , $NaLu_2Ga_3Ge_2O_{12}:Cr^{3+}$ , $NaLu_2Ga_3Ge_2O_{12}:Cr^{3+}$, $Ni^{2+}$ - b); comparison of emission spectra of $NaLu_2Ga_3Ge_2O_{12}:Cr^{3+}$, $Ni^{2+}$ with different $Ni^{2+}$ concentration at 93K -c); the influence of $Ni^{2+}$ concentration on the $LIR_1$ - d) and QY - e); luminescence decay curves of $Ni^{2+}$ ions in $NaLu_2Ga_3Ge_2O_{12}:Cr^{3+}$, $Ni^{2+}$ for different $Ni^{2+}$ ions concentration -f;) the influence of $Ni^{2+}$ ions concentration on the $\tau$ of the $^4T_2$ state of $Cr^{3+}$ ions and $\eta$ – g).

To evaluate the application potential of $NaLu_2Ga_3Ge_2O_{12}:Ni^{2+}, Cr^{3+}$ for remote temperature sensing, the emission spectra of $NaLu_2Ga_3Ge_2O_{12}:Ni^{2+}, Cr^{3+}$ were recorded as a function of temperature over the 93-553 K range (Figure 3a). The obtained results reveal pronounced temperature-induced changes in the luminescence intensities of both $Cr^{3+}$ and $Ni^{2+}$ ions. However, the thermal evolution of the two emission components differs significantly. As illustrated by the normalized luminescence map in Figure 3b, at low $Ni^{2+}$ concentrations, the $Ni^{2+}$ emission intensity initially increases with increasing temperature, followed by thermal quenching at elevated temperatures. To investigate this behavior in greater detail, the temperature dependences of the integrated $Cr^{3+}$ and $Ni^{2+}$ luminescence intensities were analyzed as a function of $Ni^{2+}$ concentration. For the $^4T_2 \rightarrow ^4A_2$ emission band of $Cr^{3+}$, the luminescence intensity decreases monotonically with increasing temperature, regardless of the $Ni^{2+}$ concentration (Figure 3c). Importantly, the rate of thermal quenching becomes progressively higher with increasing $Ni^{2+}$ content. This behavior can be attributed to the enhanced depopulation of the $Cr^{3+}$ $^4T_2$ excited state through $Cr^{3+} \rightarrow Ni^{2+}$ energy transfer. Increasing the $Ni^{2+}$ concentration reduces the average distance between the interacting $Cr^{3+}$ and $Ni^{2+}$ ions, thereby enhancing the probability of energy transfer and accelerating the temperature-induced depopulation of the $Cr^{3+}$ $^4T_2$ state. A distinctly different thermal behavior is observed for the $Ni^{2+}$ emission (Figure 3d). At low $Ni^{2+}$ concentrations, the intensity of the $^3T_2(^3F) \rightarrow ^3A_2(^3F)$ emission band initially increases with temperature, reaching approximately 150% of its value at 93 K at a temperature of around 250 K. A further increase in temperature results in a gradual decrease in the $Ni^{2+}$ luminescence intensity due to the increasing contribution of nonradiative relaxation processes. Interestingly, the magnitude of the initial thermally induced enhancement decreases systematically with increasing $Ni^{2+}$ concentration. For $Ni^{2+}$ concentrations above 0.03%, a monotonic decrease in the $Ni^{2+}$ luminescence intensity is observed throughout the entire investigated temperature range. Considering that $Cr^{3+}$→$Ni^{2+}$ energy transfer constitutes an important pathway for populating the $Ni^{2+}$ excited state, the initial temperature-induced enhancement of the $Ni^{2+}$ luminescence can be attributed to thermally assisted population of the $Ni^{2+}$ excited state through energy transfer from $Cr^{3+}$. At low $Ni^{2+}$ concentrations, increasing temperature enhances this population pathway, resulting in the observed increase in $Ni^{2+}$ emission intensity. Although increasing the $Ni^{2+}$ concentration is expected to facilitate $Cr^{3+}$→$Ni^{2+}$ energy transfer by reducing the average $Cr^{3+}$-$Ni^{2+}$ distance, it simultaneously enhances the contribution of nonradiative depopulation pathways of the $Ni^{2+}$ excited state. Consequently, at higher $Ni^{2+}$ concentrations, nonradiative relaxation progressively competes with the energy-transfer-assisted population of the emitting state, suppressing the thermally induced enhancement of $Ni^{2+}$ luminescence and eventually leading to its monotonic thermal quenching. The markedly different temperature dependences of the $Cr^{3+}$ and $Ni^{2+}$ emission intensities provide a favorable basis for ratiometric temperature sensing. Therefore, the thermal dependence of $LIR_1$ was analyzed. The obtained results demonstrate that for the sample containing 0.003% $Ni^{2+}$, $LIR_1$ increases approximately sixfold with increasing temperature, reaching its maximum at around 350 K, above which its value begins to decrease (Figure 3e). Increasing the $Ni^{2+}$ concentration progressively reduces the maximum $LIR_1$ value.

Moreover, the temperature at which the $LIR_1$ maximum is reached systematically shifts toward lower temperatures with increasing $Ni^{2+}$ concentration. These results demonstrate that the $Ni^{2+}$ content provides an effective means of controlling both the magnitude and the operating temperature range of the ratiometric thermometric response. To quantitatively evaluate the temperature dependence of $LIR_1$, the relative thermal sensitivity ($S_R$) was determined according to the following equation:

$$S_{R,T} = \frac{1}{LIR}\frac{\Delta LIR}{\Delta T}\cdot 100\% \qquad (5)$$

Consistent with the pronounced thermal variation of $LIR_1$, the highest $S_R$ value of 1.41% $K^{-1}$ is obtained for the sample containing 0.003% $Ni^{2+}$ (Figure 3f). A progressive increase in the $Ni^{2+}$ concentration results in a systematic reduction of the maximum relative thermal sensitivity, confirming that low $Ni^{2+}$ concentrations are particularly favorable for maximizing the thermometric performance of the $Cr^{3+}$/$Ni^{2+}$-based ratiometric readout.

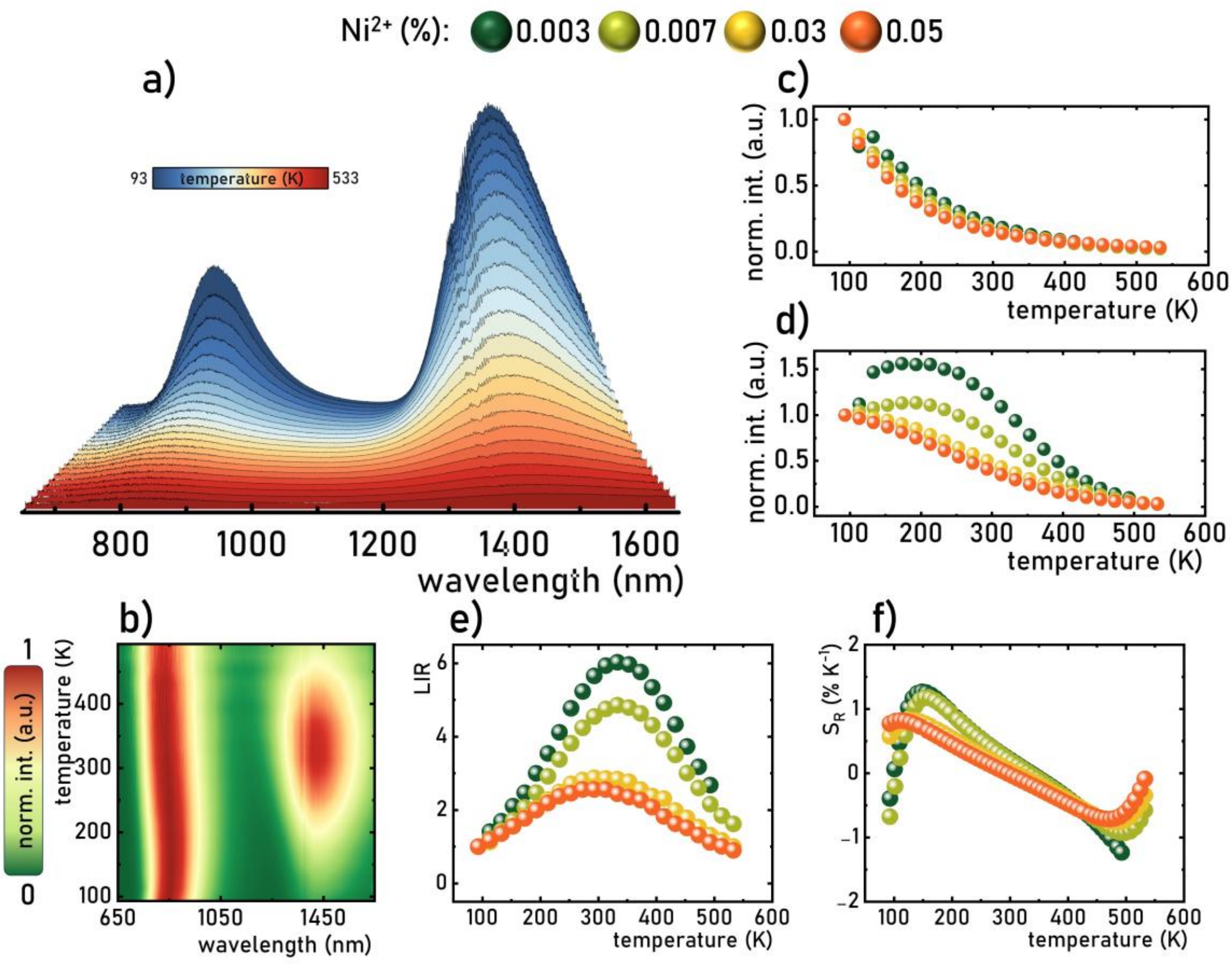

**Figure 3**. Thermal dependence of emission spectra of representative $NaLu_2Ga_3Ge_2O_{12}$:$Ni^{2+}$, $Cr^{3+}$ - a) and corresponding thermal map of normalized emission spectra - b); the influence of the temperature on the normalized emission intensity of $Cr^{3+}$ ions - c) and $Ni^{2+}$ ions -d) for different $Ni^{2+}$ ions concentrations, thermal dependence of $LIR_1$ - e) and corresponding $S_R$ - f) for different $Ni^{2+}$ ions concentrations.

Transition-metal ions are highly sensitive to variations in their local coordination environment[42,47,48]. One of the key parameters governing their spectroscopic properties is the crystal field strength experienced by the optically active ion. Since the crystal field strength strongly depends on the metal-oxygen distance, compression of the host lattice is expected to substantially modify the spectroscopic properties of both $Cr^{3+}$ and $Ni^{2+}$ ions. To investigate this effect in $NaLu_2Ga_3Ge_2O_{12}$:$Ni^{2+}$, $Cr^{3+}$, room-temperature luminescence spectra were recorded as a function of pressure (Figure 4a). As can be seen from the normalized emission spectra, increasing pressure induces a pronounced blueshift of both the $Cr^{3+}$ and $Ni^{2+}$ emission bands. Such behavior is consistent with the pressure-induced enhancement of the crystal field strength resulting from shortening of the metal-oxygen distances. According to the corresponding Tanabe-Sugano diagrams, an increase in crystal field strength raises the energy of the $^4T_2$ excited state of $Cr^{3+}$ and the $^3T_2$ state of $Ni^{2+}$, resulting in the observed shift of their emission bands toward shorter wavelengths. A detailed analysis of the $Cr^{3+}$ ions emission band associated with the $^4T_2 \rightarrow ^4A_2$ electronic transition reveals an additional pressure-induced modification of the electronic structure. At pressures above approximately 4.5 GPa, a narrow emission band associated with the $^2E \rightarrow ^4A_2$ transition becomes clearly distinguishable. The appearance of the emission band from the $^2E$ state indicates that lattice compression modifies the crystal field experienced by $Cr^{3+}$ ions sufficiently to shift the system from the weak- toward the intermediate-crystal-field regime, thereby changing the relative energies of the $^4T_2$ and $^2E$ excited states. Analysis of the pressure dependence of the emission-band maxima confirms the pronounced spectral response of both luminescent centers. The $Ni^{2+}$ emission maximum shifts from approximately 1450 nm at ambient pressure to 1300 nm at 8.83 GPa (Figure 4b). Over the corresponding pressure range, the $Cr^{3+}$ emission maximum shifts from approximately 835 nm at ambient pressure to 810 nm at 5.5 GPa (Figure 4b). At higher pressures, the intensity of the $Cr^{3+}$ $^4T_2 \rightarrow ^4A_2$ emission becomes negligible, preventing reliable determination of its spectral position. To quantify these spectral changes the absolute pressure sensitivity was calculated as follows:

$$S_{A,p} = \frac{\Delta\lambda}{\Delta p} \qquad (6)$$

where $\Delta\lambda$ represents the change of the band maxima corresponding to the change of the pressure by $\Delta p$. As it can be noticed in the case of the $Cr^{3+}$ emission band the $S_A$ decreases with pressure from $S_A$=16 nm GPa$^{-1}$ at ambient pressure to around 3.9 nm GPa$^{-1}$ at 5 GPa (Figure 4c). On the other hand much higher $S_A$ was found in the case of $Ni^{2+}$ emission band for which absolute sensitivity decreases from $S_A$ = 34 nm GPa$^{-1}$ at ambient pressure to $S_A$ = 5 nm GPa$^{-1}$ at 8.83 GPa. The comparison of the previously reported values of $S_A$ for other luminescence manometers reveals that $NaLu_2Ga_3Ge_2O_{12}:Cr^{3+}, Ni^{2+}$ is characterized by the highest up to date reported $S_A$ in the literature (Figure 4d)[12,23,24,49–58] . In addition to the spectral shifts, a pronounced redistribution of the relative $Cr^{3+}$ and $Ni^{2+}$ emission intensities is observed with increasing pressure. In particular, the $Cr^{3+}$ emission intensity decreases substantially relative to that of $Ni^{2+}$. One possible explanation for this behavior is an enhancement of the $Cr^{3+}\rightarrow Ni^{2+}$ energy-transfer efficiency under compression. The pressure-induced shortening of the average $Cr^{3+}$-$Ni^{2+}$ interionic distance may increase the interaction probability between the donor and acceptor ions, thereby promoting energy transfer $Cr^{3+}\rightarrow Ni^{2+}$ and contributing to the observed redistribution of the emission intensities. The pronounced difference in the pressure dependence of the $Cr^{3+}$ and $Ni^{2+}$ emission intensities enables ratiometric pressure sensing using the previously defined $LIR_1$ parameter (Figure 4e). The most significant pressure-induced variation of $LIR_1$ is observed below approximately 4 GPa, indicating that this pressure range is particularly favorable for ratiometric pressure determination. To quantitatively evaluate the manometric performance, the relative pressure sensitivity was calculated (Figure 4f):

$$S_{R,p} = \frac{1}{LIR}\frac{\Delta LIR}{\Delta p} \cdot 100\% \qquad (7)$$

Its maximum value reaches 22% GPa$^{-1}$ at approximately 2 GPa, demonstrating a pronounced response of $LIR_1$ to relatively small pressure variations. To evaluate the selectivity of $LIR_1$ toward pressure variations and its susceptibility to thermal interference, the thermal invariability manometric factor (*TIMF*) was determined according to the following equation:

$$TIMF = \frac{S_{R,p}}{S_{R,T}} \qquad (8)$$

The maximum *TIMF* value reaches approximately 2900 K GPa$^{-1}$ at 2 GPa. This relatively high value indicates a strong predominance of the pressure response over the temperature response of $LIR_1$ (Figure 4g). In practical terms, a temperature variation of approximately 2900 K would be required to induce a change in $LIR_1$ comparable to that produced by a pressure variation of 1 GPa. These results demonstrate that $NaLu_2Ga_3Ge_2O_{12}:Ni^{2+}$, $Cr^{3+}$ provides a highly pressure-selective ratiometric luminescence response with limited susceptibility to temperature variations, supporting its potential for reliable pressure sensing under conditions where simultaneous temperature fluctuations may occur.

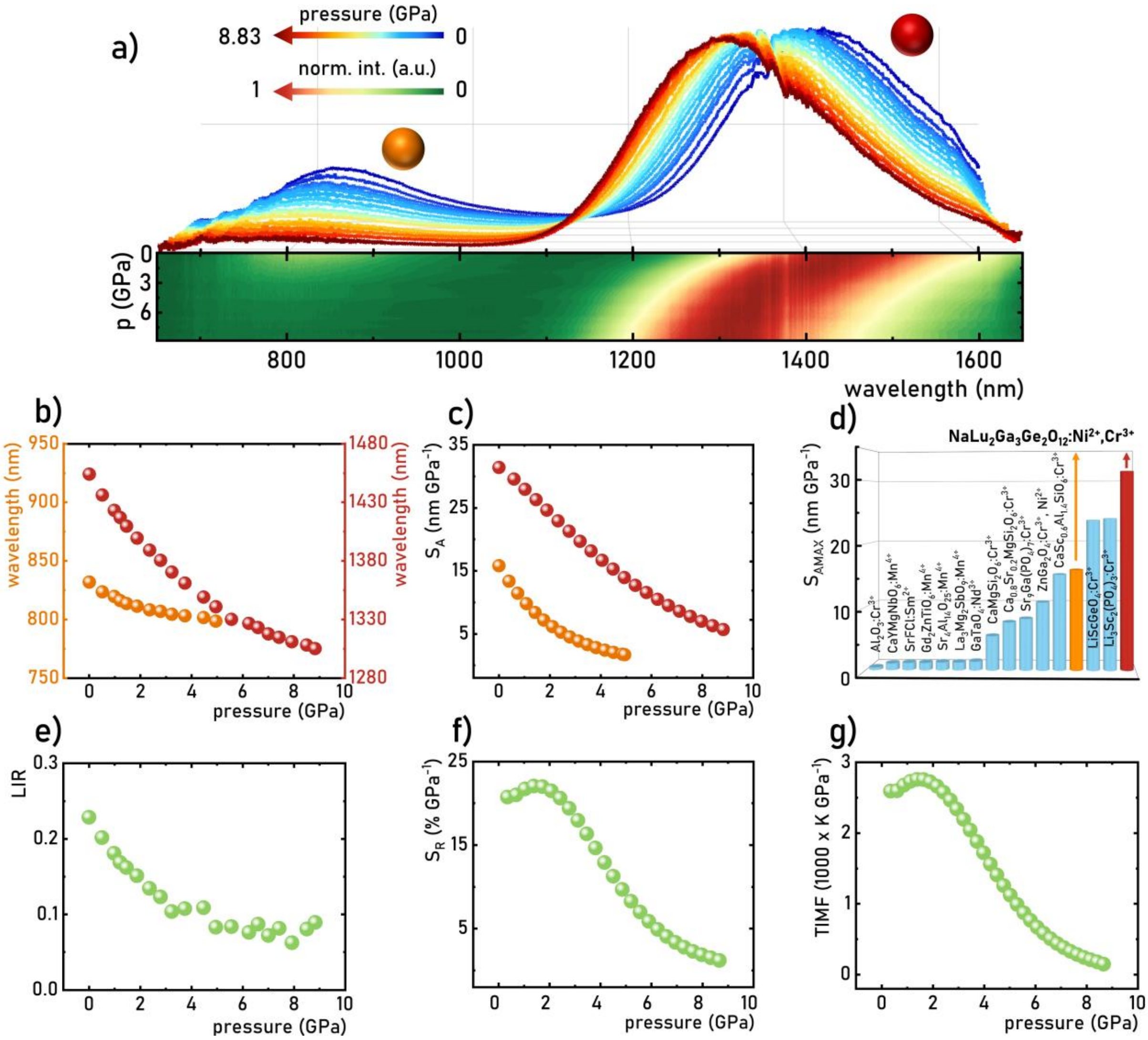

**Figure 4**. Normalized emission spectra of $NaLu_2Ga_3Ge_2O_{12}:Ni^{2+}, Cr^{3+}$ measured as a function of pressure and corresponding pressure maps of normalized emission spectra - a) the influence of the pressure on the spectral position of the emission band maxima of $Ni^{2+}$ and $Cr^{3+}$ ions - b); and corresponding $S_A$ - c) the comparison of the $S_{AMAX}$ for different luminescent manometers based on the spectral position of emission band - d); the influence of pressure on $LIR_1$ - e) and corresponding $S_R$ - f) and $TIMF$ - g).

The pressure-induced spectral shifts of the $Cr^{3+}$ and $Ni^{2+}$ emission bands can also be exploited for ratiometric pressure sensing by monitoring the luminescence intensities within two narrow spectral ranges selected for each emission band. The spectral windows indicated in Figure 5a were selected to provide an appropriate compromise between pronounced pressure-induced intensity variation and sufficiently high luminescence intensity. The general concept of this approach relies on the pressure-induced blueshift of the emission bands. As the emission maximum shifts toward shorter wavelengths, the luminescence intensity integrated over the short-wavelength side of the band increases, whereas the intensity recorded within the spectral range corresponding to its long-wavelength side decreases. Consequently, the ratio between these two spectral contributions provides a highly pressure-dependent parameter without relying on absolute luminescence intensity. An important advantage of this approach from an application perspective is its potentially low susceptibility to thermal interference. Although temperature-induced changes in the local crystal field cannot be completely excluded, their influence on the spectral positions of the $Cr^{3+}$ and $Ni^{2+}$ emission bands is expected to be substantially weaker than the changes induced by lattice compression. Therefore, spectral-shift-based ratiometric parameters should provide enhanced selectivity toward pressure variations. To verify this assumption, $LIR_2$ and $LIR_3$, corresponding to the $Ni^{2+}$ and $Cr^{3+}$ emission bands, respectively, were defined as follows:

$$LIR_2 = \frac{\int_{1200\mathrm{nm}}^{1220nm} {}^3T_2 \rightarrow {}^3A_2 d\lambda [Ni^{2+}]}{\int_{1580\mathrm{nm}}^{1600nm} {}^3T_2 \rightarrow {}^3A_2 d\lambda [Ni^{2+}]} \quad (9)$$

$$LIR_3 = \frac{\int_{780\mathrm{nm}}^{800nm} {}^4T_2 \rightarrow {}^4A_2 d\lambda [Cr^{3+}]}{\int_{830\mathrm{nm}}^{850nm} {}^4T_2 \rightarrow {}^4A_2 d\lambda [Cr^{3+}]} \quad (10)$$

The obtained results reveal a considerably stronger pressure dependence of $LIR_2$, associated with the $Ni^{2+}$ emission band, than that of $LIR_3$, derived from the $Cr^{3+}$ emission (Figure 5b). Over the investigated pressure range, $LIR_2$ increases by more than 40-fold, whereas $LIR_3$ changes by only approximately a factor of two. This behavior is consistent with the substantially larger pressure-induced spectral shift observed for the $Ni^{2+}$ emission band compared with that of $Cr^{3+}$. As a consequence, the maximum relative pressure sensitivity determined for $LIR_3$ reaches approximately 28% $GPa^{-1}$ at around 0.5 GPa (Figure 5c). At the same pressure, the maximum $S_R$ obtained for $LIR_2$ reaches 74% $GPa^{-1}$. Both values are considerably higher than that obtained using the previously discussed $LIR_1$ parameter. Importantly, the maximum sensitivities of $LIR_2$ and $LIR_3$ are reached at substantially lower pressures than the maximum sensitivity of $LIR_1$. These results demonstrate that the use of different ratiometric approaches in $NaLu_2Ga_3Ge_2O_{12}:Cr^{3+}$, $Ni^{2+}$ enables the manometric performance, including both pressure sensitivity and the optimal operating pressure range, to be tailored according to the requirements of a particular application. As discussed above, increasing pressure also induces the appearance of the narrowband $Cr^{3+}$ ${}^2E \rightarrow {}^4A_2$ emission, resulting in a pronounced increase in luminescence intensity within the corresponding spectral region. To further enhance the pressure response, this effect was combined with the pronounced pressure-induced spectral shift of the $Ni^{2+}$ emission band to define an additional ratiometric parameter, $LIR_4$:

$$LIR_4 = \frac{\int_{780\mathrm{nm}}^{800nm} {}^4T_2 \rightarrow {}^4A_2 d\lambda [Cr^{3+}]}{\int_{1580\mathrm{nm}}^{1600nm} {}^3T_2 \rightarrow {}^3A_2 d\lambda [Ni^{2+}]} \tag{11}$$

The pressure dependence of $LIR_4$ is substantially more pronounced than those observed for either $LIR_2$ or $LIR_3$ (Figure 5d). Consequently, $LIR_4$ provides a maximum relative pressure sensitivity of 158% $GPa^{-1}$ at approximately 0.2 GPa, demonstrating the significant advantage of simultaneously exploiting two pressure-dependent spectroscopic effects within a single ratiometric readout (Figure 5e). To assess the susceptibility of $LIR_4$ to temperature variations, its relative thermal sensitivity was determined analogously and used to calculate *TIMF*. A maximum *TIMF* of approximately 19,900 K $GPa^{-1}$ is obtained at 0.2 GPa (Figure 5f). This exceptionally high value demonstrates that the pressure-induced variation of $LIR_4$ strongly dominates over its temperature dependence, enabling highly selective pressure readout with minimal thermal interference. To the best of our knowledge, this represents the highest *TIMF* value reported to date for a ratiometric luminescent manometer, highlighting the exceptional potential of $NaLu_2Ga_3Ge_2O_{12}$:$Cr^{3+}$, $Ni^{2+}$ for reliable pressure sensing, particularly in the low-pressure regime (Figure 5g)[22–25,53,57–64] .

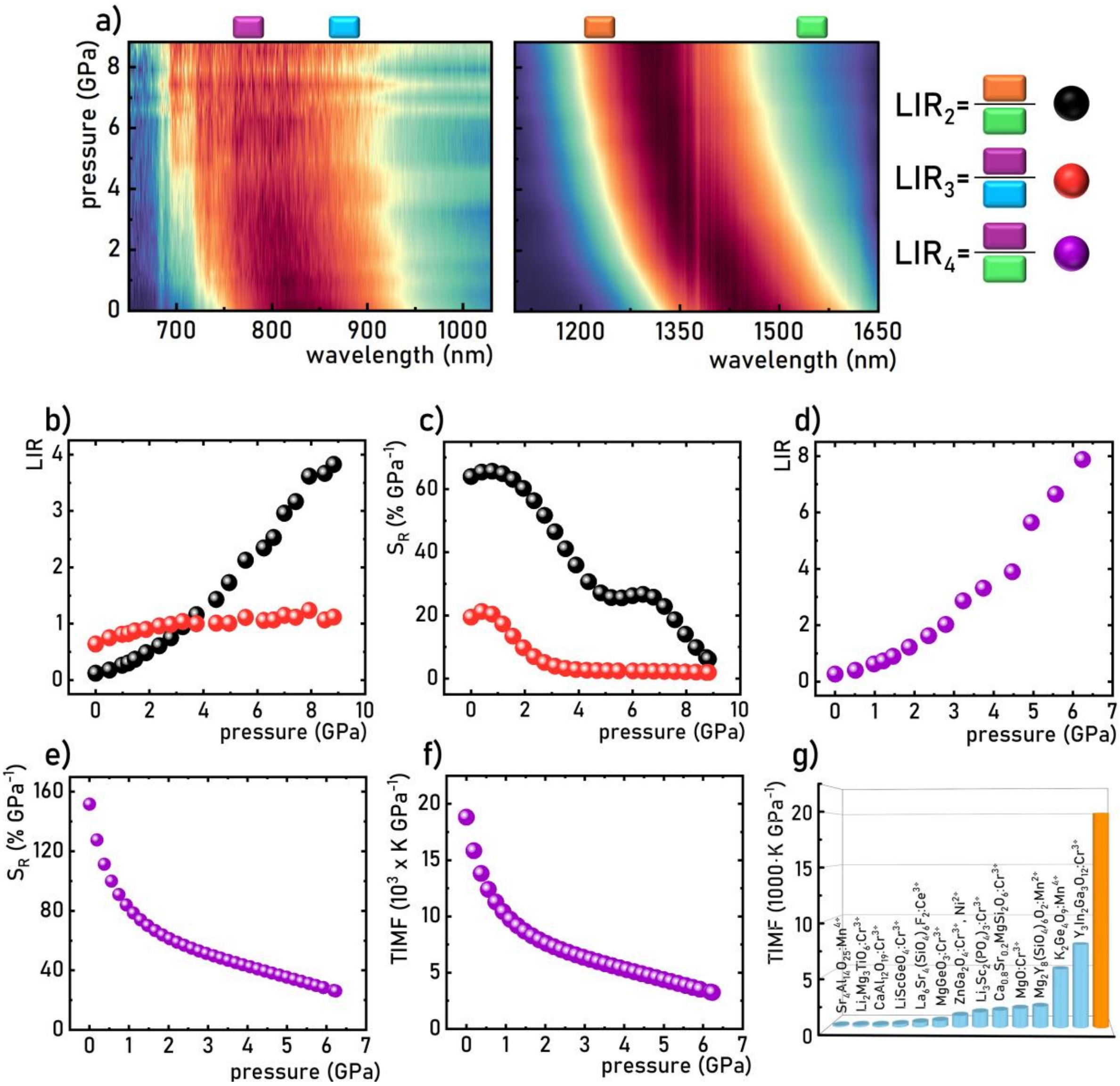

**Figure 5**. The representative normalized emission spectra of $NaLu_2Ga_3Ge_2O_{12}$:$Ni^{2+}$, $Cr^{3+}$ measured as a function of pressure ($Cr^{3+}$ and $Ni^{2+}$ emission bands normalized independently to their maximal values) with marked spectral regions used for *LIRs* calculations - a); pressure dependence of $LIR_2$ and $LIR_3$ - b) and corresponding $S_R$ - c); pressure dependence of $LIR_3$ - d) and corresponding $S_R$ - e) and *TIMF* - f); comparison of the *TIMF* parameter for different ratiometric manometers – g).

## Conclusion

In this work, the effects of temperature and pressure on the spectroscopic properties of $NaLu_2Ga_3Ge_2O_{12}:Cr^{3+}$, $Ni^{2+}$ were systematically investigated to evaluate its potential for remote pressure sensing. For this purpose, luminescence spectra were recorded as a function of dopant concentration, temperature over the range of 93-553 K, and pressure from ambient conditions up to 9 GPa. The obtained results revealed several important features that determine the sensing performance of this material. As demonstrated, co-doping $NaLu_2Ga_3Ge_2O_{12}:Ni^{2+}$ with $Cr^{3+}$ ions significantly enhanced the $Ni^{2+}$ luminescence intensity through $Cr^{3+}\rightarrow Ni^{2+}$ energy transfer. Owing to the high absorption cross-section of $Cr^{3+}$, this sensitization mechanism enabled a substantial enhancement of the $Ni^{2+}$ emission intensity. Importantly, the $Cr^{3+}\rightarrow Ni^{2+}$ energy-transfer process was found to play a crucial role in determining the manometric properties of the material. For low $Ni^{2+}$ concentrations, increasing temperature resulted in a thermally induced enhancement of the $Ni^{2+}$ luminescence, reaching approximately 150% of the intensity recorded at 93 K at around 250 K. This effect was progressively suppressed with increasing $Ni^{2+}$ concentration. The different temperature dependences of the integrated $Cr^{3+}$ and $Ni^{2+}$ emission intensities enabled the development of a ratiometric luminescence thermometer, which exhibited a maximum relative thermal sensitivity of of 1.41% $K^{-1}$ at 180 K for the sample containing 0.003% $Ni^{2+}$. Compression of $NaLu_2Ga_3Ge_2O_{12}$: $Cr^{3+}$, $Ni^{2+}$ induced considerably more pronounced changes in its spectroscopic properties. Both the $Cr^{3+}$ $^4T_2\rightarrow{}^4A_2$ emission band and the $^3T_2(^3F)\rightarrow{}^3A_2$ $(^3F)$ emission of $Ni^{2+}$ exhibited pressure-induced spectral blueshifts, accompanied by substantial changes in their relative intensities. The pressure sensitivity of the manometric parameter defined on this basis reached 22 % $GPa^{-1}$ at 2 GPa, with a corresponding TIMF of 2900 K $GPa^{-1}$ at 2 GPa. However, because this parameter remained sensitive to both pressure and temperature, an alternative manometric approach was developed based on the luminescence intensity ratio between two spectral ranges associated predominantly with the $Cr^{3+}$ and $Ni^{2+}$ emission bands. This strategy resulted in substantially enhanced relative pressure sensitivities of 28% $GPa^{-1}$ and 74% $GPa^{-1}$ at around 0.5 GPa, respectively. Furthermore, the $LIR_4$ parameter was specifically designed to maximize the opposite pressure-induced variations of the $Cr^{3+}$ and $Ni^{2+}$ emission contributions. This approach yielded an exceptionally high relative pressure sensitivity of 160 % $GPa^{-1}$ at 0.5 GPa, exceeding the performance of all other sensing strategies investigated in this work. Most importantly, in contrast to the pronounced pressure dependence of $LIR_4$, temperature variation produced practically no change in this parameter. Consequently, this approach enabled highly pressure-selective sensing with negligible thermal interference and resulted in a *TIMF* of 19 900 K $GPa^{-1}$, which, to the best of our knowledge, is the highest value reported to date. Most of all to was shown that the spectral position of the $Ni^{2+}$ ions in $NaLu_2Ga_3Ge_2O_{12}:Cr^{3+}$, $Ni^{2+}$ reveals the highest up to date reported $S_A$=32 nm $GPa^{-1}$ at ambient pressure. These results demonstrate that $NaLu_2Ga_3Ge_2O_{12}:Cr^{3+}$, $Ni^{2+}$ offers considerable potential not only for highly sensitive, remote, and temperature-invariant optical pressure sensing, but also for NIR lighting applications.

**Acknowledgment:**

Authors acknowledge support within the NAWA-MOST project

BPN/BCH/2025/1/00006/U/00001. This work is supported by National Key Research and Development Program of China (No. 2026YFE0151800), the Science Foundation of Guangxi Province (No. 2025GXNSFAA069023 and No. 2022GXNSFDA080005). Maja Szymczak gratefully acknowledges the support of the Foundation for Polish Science through the START program.